\documentclass{article}
\usepackage{amsmath}
\usepackage{graphicx}
\usepackage{amsfonts}
\usepackage{amssymb}
\usepackage{geometry}
\usepackage{float}
\usepackage{bm}
\usepackage{color}
\usepackage{epstopdf}
\usepackage{cite}

\begin{document}

\title{Three dimensional DKP oscillator in a curved Snyder space}
\author{B. Hamil \thanks{hamilbilel@gmail.com}\\
D\'{e}partement de TC de SNV, Universit\'{e} Hassiba Benbouali, Chlef,
Algeria \and M. Merad \thanks{meradm@gmail.com}\\
Facult\'{e} des Sciences Exactes, Universit\'{e} de Oum El Bouaghi, 04000
Oum El Bouaghi, Algeria \and T. Birkandan \thanks{birkandant@itu.edu.tr} \\
Department of Physics, Istanbul Technical University, 34469 Istanbul, Turkey.%
}
\date{}
\maketitle

\begin{abstract}
The Snyder-de Sitter model is an extension of the Snyder model to a de
Sitter background. It is called triply special relativity (TSR) because it
is based on three fundamental parameters: speed of light, Planck mass and
cosmological constant. In this paper, we study the three dimensional DKP
oscillator for spin zero and one in the framework of Snyder-de Sitter
algebra in momentum space. By using the technique of vector spherical
harmonics the energy spectrum and the corresponding eigenfunctions are
obtained for the both cases.

\begin{description}
\item[Keywords:] noncommutative geometry; Snyder model; curved Snyder space;
DKP oscillator.

\item[PACS numbers:] 02.40.Gh, 03.65.Ge, 03.65.Pm.
\end{description}
\end{abstract}

\section{Introduction}

The noncommutative geometry plays a crucial role in the quest for a quantum
theory of gravity. The first model of noncommutative geometry was introduced
in 1947 by Snyder \cite{1} in order to regularize the divergences that arise
in quantum field theory (QFT) over the discretization of spacetime. However,
this model did not attract much attention for many years, because of the
success of the renormalisation theory, with the exception of certain works
in the sixties \cite{2,3,4,5,6}.

After several decades, Snyder's idea was revived and it was motivated by the
development of string theory \cite{7,8} and different approaches of quantum
gravity \cite{9,10}. String theory arguments to imply a lower bound on the
localization of particles in spacetime or minimal uncertainty in position.
Snyder's model can be viewed as an example of doubly special relativity
(DSR) \cite{11,12,13,14} with one more universal constant in addition to $c$%
, the speed of light in vacuum.

The Snyder algebra is generated by the spacetime $X_{\mu }$ and momenta $%
P_{\nu }$ operators, which satisfies the commutation relations, 
\begin{equation}
\left[ X_{\mu },P_{\nu }\right] =i\hslash \left( g_{\mu \nu }+\eta P_{\mu
}P_{\nu }\right) ;\text{ \ }\left[ P_{\mu },P_{\nu }\right] =0;\text{ \ }%
\left[ X_{\mu },X_{\nu }\right] =i\hslash \eta J_{\mu \nu },
\end{equation}%
where $g_{\mu \nu }$=diag($-1, 1, 1, 1$), $\eta $ is a coupling constant of
the order of the Planck length and dimensionally $\left[ \eta \right] =\left[
momentum\right] ^{-2}$. Here, $J_{\mu \nu }=X_{\mu }P_{\nu }-X_{\nu }P_{\mu
} $ are the generators of the Lorentz symmetry. During the recent years,
there has been a growing interest in studying the properties of the Snyder
model and its dynamics from different points of view \cite{15,16,17,18}, and
more recently a construction of a scalar quantum field theory on Snyder
spacetime has been proposed \cite{19}.

Over the past few years, a large amount of effort has been devoted to
generalize the Snyder model to spacetimes of constant curvature, by
introducing a new fundamental constant $\alpha $ proportional to the
cosmological constant. The resulting model is characterized by three
invariant scales, the speed of light in vacuum, $c$, a mass $\eta $ and a
length $\alpha $, and it is called triply special relativity (TSR) or Snyder
de Sitter (SdS) model \cite{20,21,22,23,24}. To our knowledge, only a few
works have studied the properties of SdS space \cite{25,26,27,28,29,30,31,32}%
.

The Duffin-Kemmer-Petiau (DKP) equation \cite{33,34,35} is a first order
relativistic wave equation and it is used to analyze relativistic
interactions of spin-0 and spin-1 hadrons with nuclei. The DKP theories
yield results that are in better agreement with experimental data than the
Klein-Gordon (KG) theory in the analysis of deuteron-nucleus and $\alpha $%
-nucleus elastic scattering and mesons \cite{36,37}. The DKP equation is
similar to the Dirac equation, where we replace the algebra of the $\gamma
^{\mu }$-matrices by the matrices $\beta ^{\mu }$ verifying the algebra 
\begin{equation}
\beta ^{\mu }\beta ^{\nu }\beta ^{\lambda }+\beta ^{\lambda }\beta ^{\nu
}\beta ^{\mu }=g^{\mu \nu }\beta ^{\lambda }+g^{\lambda \nu }\beta ^{\mu },
\end{equation}%
where, the $\beta ^{\mu }$ are $5\times 5$ matrices in the spin-zero
representation and $10\times 10$ matrices in the spin-one representation.
The DKP theory opens new ways which allow to implement other kinds of
couplings which are not possible in the KG or Proca theories.

The DKP theory has received a great attention due to its various
applications in QCD at large and short distances \cite{38}, in the deuteron
nucleus scattering observables \cite{39}, in calculation of the
meson-nucleus optical potentials \cite{40}, in pion-nucleus scattering at
medium energies \cite{41}, in the study of covariant Hamiltonian dynamics 
\cite{42} and in different topologies such as in the non-commutative space 
\cite{43,44} and in the cosmic string space-time \cite{45}, etc.

The principal aim of this paper is to solve the 3-dimensional DKP oscillator
for spins 0 and 1 in curved Snyder space. The structure of this paper is as
follows: In Sect. 2, we briefly review a Snyder-de Sitter model and the DKP
equation. In Sect.3, we study the 3-dimensional deformed DKP oscillator for
spin 0 with Snyder-de Sitter algebra in the momentum space representation.
We obtain the exact solution and the energy spectrum for this system. In
Sect.4, the eigensolutions have been obtained for the DKP oscillator in the
case of spin 1. Sect. 5 covers our conclusions.

\section{Snyder-de Sitter algebra}

In the non-relativistic Snyder-de Sitter model, the deformed Heisenberg
algebra in 3-dimensional case is given by \cite{25,26},%
\begin{equation}
\left[ X_{i},P_{j}\right] =i\hbar \left( \delta _{ij}+\alpha X_{i}X_{j}+\eta
P_{i}P_{j}+\sqrt{\alpha \eta }\left( P_{i}X_{j}+X_{j}P_{i}\right) \right) ,
\label{3}
\end{equation}%
\begin{equation}
\left[ X_{i},P_{j}\right] =i\hbar \eta \varepsilon _{ijk}L_{k}\text{ ; \  \  \
\ }\left[ P_{i},P_{j}\right] =i\hbar \alpha \varepsilon _{ijk}L_{k}.
\end{equation}%
Here, $L_{k}$ are the components of angular momentum operator and $\alpha $
and $\eta $ are small positive parameters. In the limit $\alpha \rightarrow
0 $ one recovers the Snyder algebra, in the limit $\eta \rightarrow 0$ one
recovers the deformed Heisenberg algebra in de Sitter space endowed with
projective coordinates \cite{46}, and when $\alpha $ and $\eta $ both tend
to zero one recovers the undeformed Heisenberg algebra.

In the simple case in which $\left \langle X_{i}\right \rangle
=\left
\langle P_{i}\right \rangle =0$, the uncertainty relation associated
with (\ref{3}) can be written as%
\begin{equation}
\left( \Delta X_{i}\right) \left( \Delta P_{j}\right) \geqslant \frac{\hbar 
}{2}\left( \delta _{ij}+\alpha \left( \Delta X_{i}\right) \left( \Delta
X_{j}\right) +\eta \left( \Delta P_{i}\right) \left( \Delta P_{j}\right) +%
\sqrt{\alpha \eta }\left( \left( \Delta P_{i}\right) \left( \Delta
X_{j}\right) +\left( \Delta X_{j}\right) \left( \Delta P_{i}\right) \right)
\right) .  \label{5}
\end{equation}%
In the particular case when $i=j$, the uncertainty relation (\ref{5}) reduce
to%
\begin{equation}
\left( \Delta X\right) \left( \Delta P\right) \geqslant \frac{\hbar }{%
2\left( 1+\hbar \sqrt{\alpha \eta }\right) }\left( 1+\alpha \left( \Delta
X\right) ^{2}+\eta \left( \Delta P\right) ^{2}\right) .
\end{equation}%
The uncertainty relation above implies the appearance of minimal uncertainty
in position as well as in momentum given by 
\begin{equation}
\left( \Delta X\right) _{\min }=\hbar \sqrt{\frac{\eta }{1+2\hbar \sqrt{%
\alpha \eta }}};\text{ \  \  \  \  \ }\left( \Delta P\right) _{\min }=\hbar 
\sqrt{\frac{\alpha }{1+2\hbar \sqrt{\alpha \eta }}}.
\end{equation}%
If $\alpha ,$ $\eta <0$, then no minimal uncertainty in position and in
momentum arises. In the momentum representation, the operators $X_{i}$ and $%
P_{i}$ can be written as 
\begin{equation}
X_{j}=i\hbar \sqrt{1-\eta p^{2}}\frac{\partial }{\partial p_{j}}+\sqrt{\frac{%
\eta }{\alpha }}\lambda \frac{p_{j}}{\sqrt{1-\eta p^{2}}},  \label{8}
\end{equation}%
\begin{equation}
P_{j}=-i\hbar \sqrt{\frac{\alpha }{\eta }}\sqrt{1-\eta p^{2}}\frac{\partial 
}{\partial p_{j}}+\left( 1-\lambda \right) \frac{p_{i}}{\sqrt{1-\eta p^{2}}},
\label{9}
\end{equation}%
where $p^{2}=\sum_{i=1}^{3}p_{i}^{2},$ $\lambda $ is an arbitrary parameter
and $p$ varies in the domain $\left] -\frac{1}{\eta };\frac{1}{\eta }\right[
.$ In order for the operators $X_{i}$ and $P_{i}$ to be symmetric, the
scalar product must be defined as%
\begin{equation}
\left \langle \phi \right. \left \vert \psi \right \rangle =\int_{-\frac{1}{%
\eta }}^{\frac{1}{\eta }}\frac{d\overrightarrow{p}}{\sqrt{1-\eta p^{2}}}\phi
^{\ast }\left( p\right) \psi \left( p\right) .
\end{equation}%
On the other hand, the DKP equation for a free boson is given by 
\begin{equation}
\left( c\overrightarrow{\beta }.\overrightarrow{p}+mc^{2}\right) \psi =\beta
_{0}E\psi .
\end{equation}%
For the case of spin-zero, the explicit expressions of the five-dimensional $%
\beta ^{\mu }$-matrices are 
\begin{equation}
\beta ^{0}=\left( 
\begin{array}{cc}
\theta & \mathbf{0} \\ 
\bar{0}_{T} & 0%
\end{array}%
\right) ;\text{ }\beta ^{i}=\left( 
\begin{array}{cc}
\tilde{0} & \rho _{i} \\ 
-\rho _{i}^{T} & \mathbf{0}%
\end{array}%
\right) ,
\end{equation}%
where 
\begin{equation}
\theta =\left( 
\begin{array}{cc}
0 & 1 \\ 
1 & 0%
\end{array}%
\right) ;\text{ }\rho _{1}=\left( 
\begin{array}{ccc}
-1 & 0 & 0 \\ 
0 & 0 & 0%
\end{array}%
\right) ;\text{ }\rho _{2}=\left( 
\begin{array}{ccc}
0 & -1 & 0 \\ 
0 & 0 & 0%
\end{array}%
\right) ;\text{ }\rho _{3}=\left( 
\begin{array}{ccc}
0 & 0 & -1 \\ 
0 & 0 & 0%
\end{array}%
\right) .
\end{equation}%
Here, $\mathbf{0,}$ $\tilde{0}$ and $\bar{0}$ are $3\times 3\mathbf{,}$ $%
2\times 2$ and $2\times 3$ zero matrices, respectively. For the case of
spin-one, $\beta ^{\mu }$ are $10\times 10$ matrices expressed as 
\begin{equation}
\beta ^{0}=\left( 
\begin{array}{cccc}
0 & \bar{0} & \bar{0} & \bar{0} \\ 
\bar{0}^{T} & \mathbf{0} & \mathbf{1} & \mathbf{0} \\ 
\bar{0}^{T} & \mathbf{1} & \mathbf{0} & \mathbf{0} \\ 
\bar{0}^{T} & \mathbf{0} & \mathbf{0} & \mathbf{0}%
\end{array}%
\right) ;\text{ }\beta _{j}=\left( 
\begin{array}{cccc}
0 & \bar{0} & e_{j} & \bar{0} \\ 
\bar{0}^{T} & \mathbf{0} & \mathbf{0} & -iS_{j} \\ 
-e_{j}^{T} & \mathbf{0} & \mathbf{0} & \mathbf{0} \\ 
\bar{0}^{T} & -iS_{j} & \mathbf{0} & \mathbf{0}%
\end{array}%
\right) ,
\end{equation}%
where the matrices $S_{j}$ are the usual $3\times 3$ spin-1 matrices, and $%
\mathbf{1}$ is $3\times 3$ unity matrix, the matrices $\bar{0}$ and $e_{j}$
are given as 
\begin{equation}
\bar{0}=\left( 
\begin{array}{ccc}
0 & 0 & 0%
\end{array}%
\right) ;\text{ \ }e_{1}=\left( 
\begin{array}{ccc}
1 & 0 & 0%
\end{array}%
\right) ;\text{ \ }e_{2}=\left( 
\begin{array}{ccc}
0 & 1 & 0%
\end{array}%
\right) ;\text{ \  \ }e_{3}=\left( 
\begin{array}{ccc}
0 & 0 & 1%
\end{array}%
\right) .
\end{equation}

\section{ Scalar DKP oscillator}

In this section, we study the dynamics of a spin-zero particle in curved
Snyder space. The DKP oscillator system is introduced by the substitution, 
\begin{equation}
\overrightarrow{P}\rightarrow \overrightarrow{P}-i\eta ^{0}m\omega 
\overrightarrow{X},
\end{equation}%
with $\omega $ being the oscillator frequency and $\eta ^{0}=2\left( \beta
^{0}\right) ^{2}-\mathbf{1}.$ Thus, the DKP equation for the DKP oscillator
system is%
\begin{equation}
\left[ c\overrightarrow{\beta }\left( \overrightarrow{P}-i\eta ^{0}m\omega 
\overrightarrow{X}\right) +mc^{2}\right] \psi =\beta _{0}E\psi .  \label{17}
\end{equation}%
The wave function $\psi $ has 5-components which can be written as 
\begin{equation}
\psi =\left( 
\begin{array}{c}
\Psi _{upper} \\ 
\Psi _{lower}%
\end{array}%
\right) \text{ with \ }\Psi _{upper}=\left( 
\begin{array}{c}
\digamma _{1} \\ 
\digamma _{2}%
\end{array}%
\right) \text{ and \  \ }\Psi _{lower}=\left( 
\begin{array}{c}
\digamma _{3} \\ 
\digamma _{4} \\ 
\digamma _{5}%
\end{array}%
\right) ,
\end{equation}%
and this 5-component wavefunction is simultaneously an eigenfunction of $%
J^{2}$ and $J_{z}$,%
\begin{eqnarray}
\hat{J}^{2}\psi &=&\left( 
\begin{array}{c}
\hat{L}^{2}\Psi _{upper} \\ 
\left( \hat{L}+\hat{S}\right) ^{2}\Psi _{lower}%
\end{array}%
\right) =J\left( J+1\right) \left( 
\begin{array}{c}
\Psi _{upper} \\ 
\Psi _{lower}%
\end{array}%
\right) , \\
\hat{J}_{z}\psi &=&\left( 
\begin{array}{c}
\hat{L}_{z}\Psi _{upper} \\ 
\left( \hat{L}_{z}+\hat{S}_{z}\right) \Psi _{lower}%
\end{array}%
\right) =M\left( 
\begin{array}{c}
\Psi _{upper} \\ 
\Psi _{lower}%
\end{array}%
\right) .
\end{eqnarray}%
For the DKP oscillator, the general solution is considered as%
\begin{equation}
\psi _{nJM}=\left( 
\begin{array}{c}
f_{nJ}\left( p\right) Y_{JM}\left( p_{\Omega }\right) \\ 
g_{nJ}\left( p\right) Y_{JM}\left( p_{\Omega }\right) \\ 
i\sum \limits_{L}h_{nJL}\left( p\right) Y_{JL1}^{M}\left( p_{\Omega }\right)%
\end{array}%
\right) ,  \label{21}
\end{equation}%
where the spherical harmonics $Y_{JM}\left( p_{\Omega }\right) $ are of the
order $J$ and, $f_{nJ}\left( p\right) $, $g_{nJ}\left( p\right) $, $%
h_{nJL}\left( p\right) $ are radial wave functions. Here, 
\begin{equation}
Y_{JL1}^{M}\left( p_{\Omega }\right) =\sum \limits_{\lambda ,\mu }\left
\langle JM\right \vert \left. L\lambda 1\mu \right \rangle Y_{L\lambda
}\left( p_{\Omega }\right) \chi _{1\mu },
\end{equation}%
are the normalized vector spherical harmonics. To get a solution of the Eq. (%
\ref{17}) we substitute (\ref{21}) into (\ref{17}) to get the coupled
equations as 
\begin{equation}
mc^{2}f_{nJ}Y_{JM}+c\left( -\hbar \theta \sqrt{\frac{\alpha }{\eta }}\sqrt{%
1-\eta p^{2}}\overrightarrow{\nabla }-i\left( 1-\lambda \theta \right) \frac{%
\overrightarrow{p}}{\sqrt{1-\eta p^{2}}}\right) \sum
\limits_{L}h_{nJL}Y_{JL1}^{M}=Eg_{nJ}Y_{JM},
\end{equation}%
\begin{equation}
mc^{2}g_{nJ}=Ef_{nJ},
\end{equation}%
\begin{equation}
c\left( -i\hbar \theta ^{\ast }\sqrt{\frac{\alpha }{\beta }}\sqrt{1-\eta
p^{2}}\overrightarrow{\nabla }+\left( 1-\lambda \theta ^{\ast }\right) \frac{%
\overrightarrow{p}}{\sqrt{1-\eta p^{2}}}\right) f_{nJ}Y_{JM}+imc^{2}\sum
\limits_{L}h_{nJL}Y_{JL1}^{M}=0,
\end{equation}%
where%
\begin{equation}
\theta =1-\frac{im\omega \sqrt{\eta }}{\sqrt{\alpha }}.
\end{equation}%
By using the properties of vector spherical harmonics \cite{47,48,49}, one
obtains the following coupled differential equations, 
\begin{equation}
mc^{2}F_{0}-\hbar c\theta \sqrt{\frac{\alpha }{\eta }}\sqrt{1-\eta p^{2}}%
\left[ 
\begin{array}{c}
-\xi _{J}\left( \frac{d}{dp}+\frac{J+1}{p}+i\sqrt{\frac{\eta }{\alpha }}%
\left( \frac{1-\lambda \theta }{\hbar \theta }\right) \frac{p}{1-\eta p^{2}}%
\right) H_{+1} \\ 
+\zeta _{J}\left( \frac{d}{dp}-\frac{J}{p}+i\sqrt{\frac{\eta }{\alpha }}%
\left( \frac{1-\lambda \theta }{\hbar \theta }\right) \frac{p}{1-\eta p^{2}}%
\right) H_{-1}%
\end{array}%
\right] =EG_{0},  \label{27}
\end{equation}%
\begin{equation}
mc^{2}G_{0}=EF_{0},  \label{28}
\end{equation}%
\begin{equation}
\hbar c\theta ^{\ast }\sqrt{\frac{\alpha }{\eta }}\sqrt{1-\eta p^{2}}\xi
_{J}\left( \frac{d}{dp}-\frac{J+1}{p}+i\sqrt{\frac{\eta }{\alpha }}\frac{%
\left( 1-\lambda \theta ^{\ast }\right) }{\hbar \theta ^{\ast }}\frac{p}{%
1-\eta p^{2}}\right) F_{0}+mc^{2}H_{+1}=0,  \label{29}
\end{equation}%
\begin{equation}
-\hbar c\theta ^{\ast }\sqrt{\frac{\alpha }{\eta }}\sqrt{1-\eta p^{2}}\zeta
_{J}\left( \frac{d}{dp}+\frac{J}{p}+i\sqrt{\frac{\eta }{\alpha }}\frac{%
\left( 1-\lambda \theta ^{\ast }\right) }{\hbar \theta ^{\ast }}\frac{p}{%
1-\eta p^{2}}\right) F_{0}+mc^{2}H_{-1}=0,  \label{30}
\end{equation}%
with definitions 
\begin{equation}
\xi _{J}=\sqrt{\frac{J+1}{2J+1}};\text{ }\zeta _{J}=\sqrt{\frac{J}{2J+1}};%
\text{ \ }f_{nJ}=\frac{F_{0}}{r};\text{ \ }g_{nJ}=\frac{G_{0}}{r};\text{ \ }%
h_{nJ,J\pm 1}=\frac{H_{\pm 1}}{r}.
\end{equation}%
Combining Eqs. (\ref{27}), (\ref{28}) and (\ref{29}), (\ref{30}), the
component $F_{0}$ can be written as%
\begin{eqnarray}
&&\bigg[\left( 1-\eta p^{2}\right) \frac{d^{2}}{dp^{2}}-\eta p\frac{d}{dp}+%
\frac{i}{\hbar }\sqrt{\frac{\eta }{\alpha }}\left[ \frac{1-\lambda \theta
^{\ast }}{\theta ^{\ast }}+\frac{1-\lambda \theta }{\theta }\right] p\frac{d%
}{dp}-\frac{J\left( J+1\right) }{p^{2}}\left( 1-\eta p^{2}\right) -\frac{%
\eta }{\alpha }\frac{\left( 1-\lambda \theta \right) \left( 1-\lambda \theta
^{\ast }\right) }{\hbar ^{2}\theta \theta ^{\ast }}\frac{p^{2}}{\left(
1-\eta p^{2}\right) }  \notag \\
&&+i\sqrt{\frac{\eta }{\alpha }}\frac{\left( 1-\lambda \theta ^{\ast
}\right) }{\hbar \theta ^{\ast }}\frac{1}{\left( 1-\eta p^{2}\right) }+\eta
+i\sqrt{\frac{\eta }{\alpha }}\left[ \frac{1-\lambda \theta ^{\ast }}{\hbar
\theta ^{\ast }}-\frac{1-\lambda \theta }{\hbar \theta }\right] +\frac{\eta
\left( E^{2}-m^{2}c^{4}\right) }{\alpha \hbar ^{2}c^{2}\theta \theta ^{\ast }%
}\bigg]F_{0}=0.  \label{32}
\end{eqnarray}%
%
%
%
%
%
%
%
%
%
In order to simplify the differential equation above, we remove the term $%
\left( \frac{1-\lambda \theta ^{\ast }}{\theta ^{\ast }}+\frac{1-\lambda
\theta }{\theta }\right) $ by setting 
\begin{equation}
\lambda =\frac{1}{\theta \theta ^{\ast }}=\frac{\alpha }{\alpha +\eta
m^{2}\omega ^{2}},
\end{equation}%
in Eq.(\ref{32}). With this choice, the differential equation for $F_{0}$
can be reduced to the following form: 
\begin{equation}
\left[ \frac{d^{2}}{dp^{2}}-\frac{\eta p}{\left( 1-\eta p^{2}\right) }\frac{d%
}{dp}-\frac{J\left( J+1\right) }{p^{2}}-\frac{\frac{\eta \mathcal{M}\Omega }{%
\hbar }\left( \frac{\mathcal{M}\Omega }{\hbar }-1\right) }{\left( 1-\eta
p^{2}\right) ^{2}}+\frac{\eta \left( \frac{2\mathcal{M}\Omega }{\hbar }%
+1\right) +\frac{\eta \left( \mathcal{E}^{2}-\mathcal{M}^{2}c^{4}+\mathcal{M}%
^{2}\Omega ^{2}c^{2}\right) }{\hbar ^{2}c^{2}}}{\left( 1-\eta p^{2}\right) }%
\right] F_{0}=0,  \label{34}
\end{equation}%
where the parameters $\mathcal{E}$, $\mathcal{M}$ and $\Omega $ are defined
as%
\begin{equation*}
\mathcal{E}=\frac{E}{\sqrt{\alpha +\eta m^{2}\omega ^{2}}};\text{ \  \ }%
\mathcal{M}=\frac{m}{\sqrt{\alpha +\eta m^{2}\omega ^{2}}};\text{ \ }\Omega =%
\frac{\omega }{\sqrt{\alpha +\eta m^{2}\omega ^{2}}}.
\end{equation*}%
In order to transform Eq. (\ref{34}) into a class of known differential
equations, we introduce the following change of variable 
\begin{equation}
\rho =\eta p^{2},
\end{equation}%
which casts Eq. (\ref{34}) into the following form, 
\begin{equation}
\left[ \rho \left( 1-\rho \right) \frac{d^{2}}{d\rho ^{2}}+\left( \frac{1}{2}%
-\rho \right) \frac{d}{d\rho }-\frac{J\left( J+1\right) }{4\rho }-\frac{%
\frac{\mathcal{M}\Omega }{4\hbar }\left( \frac{\mathcal{M}\Omega }{\hbar }%
-1\right) }{\left( 1-\rho \right) }+\frac{1}{4}\left( \frac{\mathcal{E}^{2}}{%
\hbar ^{2}c^{2}}-\frac{\mathcal{M}^{2}c^{2}}{\hbar ^{2}}+\frac{\mathcal{M}%
^{2}\Omega ^{2}}{\hbar ^{2}}+\frac{2\hbar \mathcal{M}\Omega }{\hbar ^{2}}%
+1+J\left( J+1\right) \right) \right] F_{0}=0.
\end{equation}%
We note that this equation has three regular singular points at $\rho
=0;1;\infty $. In order to rewrite this equation in the form of a known
differential equation, we make the transformation, 
\begin{equation}
F_{0}=\left( 1-\rho \right) ^{\vartheta }\rho ^{\ell }\Xi _{nJ},
\end{equation}%
then, the differential equation above is reduced to the hypergeometric
equation of the form 
\begin{equation}
\left[ \rho \left( 1-\rho \right) \frac{d^{2}}{d\rho ^{2}}+\left( \frac{3}{2}%
+J-\rho \left( 2+J+\frac{\mathcal{M}\Omega }{\hbar }\right) \right) \frac{d}{%
d\rho }+\frac{1}{4}\left( \frac{\mathcal{E}^{2}}{\hbar ^{2}c^{2}}-\frac{%
\mathcal{M}^{2}c^{2}}{\hbar ^{2}}-\left( \frac{2\mathcal{M}\Omega }{\hbar }%
+1\right) J\right) \right] F_{0}=0.  \label{38}
\end{equation}%
where%
\begin{equation}
\vartheta =\frac{\mathcal{M}\Omega }{2\hbar };\text{ \  \  \  \  \ }\ell =\frac{%
J+1}{2}.
\end{equation}%
The regular solution of the differential equation (\ref{38}) at origin $\rho
=0$ can be given in terms of the hypergeometric functions as 
\begin{equation}
\Xi =\mathbf{F}\left( a,b,J+\frac{3}{2};\rho \right) ,
\end{equation}%
whose parameters are given by 
\begin{eqnarray}
a &=&\frac{\mathcal{M}\Omega }{2\hbar }+\frac{J+1}{2}+\frac{1}{2}\sqrt{%
J\left( J+1\right) +\frac{\mathcal{E}^{2}-\mathcal{M}^{2}c^{4}}{\hbar
^{2}c^{2}}+\frac{\mathcal{M}^{2}\Omega ^{2}}{\hbar ^{2}}+\frac{2\mathcal{M}%
\Omega }{\hbar }+1}, \\
b &=&\frac{\mathcal{M}\Omega }{2\hbar }+\frac{J+1}{2}-\frac{1}{2}\sqrt{%
J\left( J+1\right) +\frac{\mathcal{E}^{2}-\mathcal{M}^{2}c^{4}}{c^{2}\hbar
^{2}}+\frac{\mathcal{M}^{2}\Omega ^{2}}{\hbar ^{2}}+\frac{2\mathcal{M}\Omega 
}{\hbar }+1}.
\end{eqnarray}%
To obtain the energy spectrum, we can use the fact that the hypergeometric
function becomes a polynomial if 
\begin{equation}
b=-n\text{ \ or \  \ }a=-n;\text{ \  \  \  \ }n=0,1,2...  \label{43}
\end{equation}%
From (\ref{43}), the energy eigenvalues of the system can be obtained as 
\begin{equation}
E_{n,J}=\pm c\sqrt{\allowbreak 2\hbar m\omega \left( 2n+J\right)
+\allowbreak m^{2}c^{2}+\left[ 4n^{2}+4n\left( J+1\right) +J\right] \hbar
^{2}\left( \alpha +\eta m^{2}\omega ^{2}\right) },
\end{equation}%
and the corresponding eigenvalues can be written in terms of the Jacobi
polynomials, namely 
\begin{equation}
F_{0}=\mathcal{N}\text{ }\eta ^{\frac{J+1}{2}}p^{J+1}\left( 1-\eta
p^{2}\right) ^{\frac{\mathcal{M}\Omega }{2\hbar }}\mathbf{P}_{n}^{\left( 
\frac{\mathcal{M}\Omega }{\hbar }-\frac{1}{2},J+\frac{1}{2}\right) }\left(
2\eta p^{2}-1\right) ,
\end{equation}%
where $\mathcal{N}$ is a normalization constant. It should be noted that due
to the modification of the Heisenberg algebra, the energy spectrum of our
system contains an additional correction term and its value increases with
the deformation parameters $\left( \alpha ,\eta \right) $. In addition, the
energy levels depend on the square of the quantum number $n$, which explains
the confinement in the high energy sector.

In the limit $\alpha \rightarrow 0$ , we obtain the energy spectrum in the
presence of minimal uncertainty in position as 
\begin{equation}
E_{n,J}=\pm c\sqrt{\allowbreak 2\hbar m\omega \left( 2n+J\right)
+\allowbreak m^{2}c^{2}+\left[ 4n^{2}+4n\left( J+1\right) +J\right] \hbar
^{2}\eta m^{2}\omega ^{2}},
\end{equation}
whereas for $\eta \rightarrow 0$, we obtain the energy spectrum in the
presence of minimal uncertainty in momentum as 
\begin{equation}
E_{n,J}=\pm c\sqrt{\allowbreak 2\hbar m\omega \left( 2n+J\right)
+\allowbreak m^{2}c^{2}+\left[ 4n^{2}+4n\left( J+1\right) +J\right] \hbar
^{2}\alpha },
\end{equation}
and for $\alpha ,$ $\eta <0,$ the energy spectrum becomes 
\begin{equation}
\frac{E_{n,J}^{2}}{c^{2}}=2\hbar m\omega \left( 2n+J\right) +\allowbreak
m^{2}c^{2}-\left[ 4n^{2}+4n\left( J+1\right) +J\right] \hbar ^{2}\left(
\alpha +\eta m^{2}\omega ^{2}\right).  \label{48}
\end{equation}
As $n$ increases, the energy spectrum (\ref{48}) becomes negative. In order
to preserve the bound $E_{n,J}^{2}\geqslant 0$ , one must impose an upper
bound $\allowbreak $on the allowed values of $n$.

By using the Jacobi polynomial property \cite{50}, 
\begin{equation}
\frac{d}{dx}\mathbf{P}_{n}^{\left( a,b\right) }\left( x\right) =\frac{n+a+b+1%
}{2}\mathbf{P}_{n-1}^{\left( a+1,b+1\right) }\left( x\right) ,
\end{equation}%
we can derive all of the spinor components as 
\begin{equation}
G_{0}=\mathcal{N}\frac{E\eta ^{\frac{J+1}{2}}}{mc^{2}}p^{J+1}\left( 1-\eta
p^{2}\right) ^{\frac{\mathcal{M}\Omega }{2\hbar }}\mathbf{P}_{n}^{\left( 
\frac{\mathcal{M}\Omega }{\hbar }-\frac{1}{2},J+\frac{1}{2}\right) }\left(
2\eta p^{2}-1\right) ,
\end{equation}%
\begin{eqnarray}
H_{-1} &=&\mathcal{N}\frac{\hbar c\sqrt{\alpha }\theta ^{\ast }\zeta
_{J}\eta ^{\frac{J}{2}}}{mc^{2}}p^{J}\left( 1-\eta p^{2}\right) ^{\frac{%
\mathcal{M}\Omega }{2\hbar }+\frac{1}{2}}\times  \notag \\
&&\left[ \left( 2J+1-\frac{\frac{\mathcal{M}\Omega }{\hbar }\eta p^{2}}{%
\left( 1-\eta p^{2}\right) }+i\sqrt{\frac{\eta }{\alpha }}\frac{\left(
1-\lambda \theta ^{\ast }\right) }{\hbar \theta ^{\ast }}p^{2}\left( 1-\eta
p^{2}\right) \right) \mathbf{P}_{n}^{\left( \frac{\mathcal{M}\Omega }{\hbar }%
-\frac{1}{2},J+\frac{1}{2}\right) }\left( 2\eta p^{2}-1\right) \right. 
\notag \\
&&\left. +2\eta \left( n+\frac{\mathcal{M}\Omega }{\hbar }+J+1\right) p^{2}%
\mathbf{P}_{n-1}^{\left( \frac{\mathcal{M}\Omega }{\hbar }+\frac{1}{2},J+%
\frac{3}{2}\right) }\left( 2\eta p^{2}-1\right) \right] ,
\end{eqnarray}%
\begin{eqnarray}
H_{+1} &=&-\mathcal{N}\frac{\hbar c\theta ^{\ast }\sqrt{\alpha }\xi _{J}\eta
^{\frac{J}{2}}}{mc^{2}}\text{ }p^{J+2}\left( 1-\eta p^{2}\right) ^{\frac{%
\mathcal{M}\Omega }{2\hbar }+\frac{1}{2}}\times  \notag \\
&&\left[ \left( i\sqrt{\frac{\eta }{\alpha }}\frac{\left( 1-\lambda \theta
^{\ast }\right) }{\hbar \theta ^{\ast }}\left( 1-\eta p^{2}\right) -\frac{%
\frac{\mathcal{M}\Omega }{\hbar }\eta }{\left( 1-\eta p^{2}\right) }\right) 
\mathbf{P}_{n}^{\left( \frac{\mathcal{M}\Omega }{\hbar }-\frac{1}{2},J+\frac{%
1}{2}\right) }\left( 2\eta p^{2}-1\right) \right.  \notag \\
&&\left. +2\eta \left( n+\frac{\mathcal{M}\Omega }{\hbar }+J+1\right) 
\mathbf{P}_{n-1}^{\left( \frac{\mathcal{M}\Omega }{\hbar }+\frac{1}{2},J+%
\frac{3}{2}\right) }\left( 2\eta p^{2}-1\right) \right] .
\end{eqnarray}

\section{Vector DKP oscillator}

In this section we will study the spin-one DKP oscillator in 3-dimensions.
The wave function $\psi $ possesses 10-components and it can be expressed as 
\begin{equation}
\psi _{nJM}=\left( 
\begin{array}{c}
i\phi _{nJ}Y_{JM} \\ 
\sum \limits_{L}f_{nJL}Y_{JL1}^{M} \\ 
\sum \limits_{L}g_{nJL}Y_{JL1}^{M} \\ 
\sum \limits_{L}h_{nJL}Y_{JL1}^{M}%
\end{array}%
\right) .  \label{53}
\end{equation}%
Following the procedure used in \cite{47,48,49}, we substitute (\ref{53}) in
(\ref{17}). We obtain ten equations which reduce to the two classes
associated with the $\left( -1\right) ^{J}$ and $\left( -1\right) ^{J+1}$
parities. The $\left( -1\right) ^{J}$ solutions correspond to the
natural-parity or magnetic-like states and $\left( -1\right) ^{J+1}$
solutions correspond to the natural-parity or electric-like states. For the
classes of $\left( -1\right) ^{J}$ parity, the relevant differential
equations are%
\begin{equation}
c\left( \overrightarrow{P}+im\omega \overrightarrow{X}\right) \times \left[
h_{nJ+1}Y_{JJ+1}^{M}+h_{nJ-1}Y_{JJ-1}^{M}\right]
+mc^{2}f_{nJ}Y_{JJL}^{M}=Eg_{nJ}Y_{JJL}^{M},
\end{equation}%
\begin{equation}
mc^{2}g_{nJ}=Ef_{nJ},
\end{equation}%
\begin{equation}
c\left( \overrightarrow{P}-im\omega \overrightarrow{X}\right) \times
f_{nJ}Y_{JJL}^{M}+mc^{2}\left[ h_{nJ+1}Y_{JJ+1}^{M}+h_{nJ-1}Y_{JJ-1}^{M}%
\right] =0.\text{\ }
\end{equation}%
Using the properties of the vector spherical harmonics, and substituting the
expressions of $X_{i}$ as given in eq. (\ref{8}) and $P_{i}$ as given in eq.
(\ref{9}), we obtain the following coupled system, 
\begin{equation}
\hbar c\theta \sqrt{\frac{\alpha }{\eta }}\sqrt{1-\eta p^{2}}\left[ 
\begin{array}{c}
\zeta _{J}\left( \frac{d}{dp}+\frac{J+1}{p}+i\sqrt{\frac{\eta }{\alpha }}%
\frac{\left( 1-\theta \lambda \right) }{\hbar \theta }\frac{p}{1-\eta p^{2}}%
\right) H_{+1}+ \\ 
\xi _{J}\left( \frac{d}{dp}-\frac{J}{p}+i\sqrt{\frac{\eta }{\alpha }}\frac{%
\left( 1-\theta \lambda \right) }{\hbar \theta }\frac{p}{1-\eta p^{2}}%
\right) H_{-1}%
\end{array}%
\right] +mc^{2}F_{0}=EG_{0},  \label{57}
\end{equation}%
\begin{equation}
EF_{0}=mc^{2}G_{0},  \label{58}
\end{equation}%
\begin{equation}
\hbar c\theta ^{\ast }\sqrt{\frac{\alpha }{\eta }}\sqrt{1-\eta p^{2}}\zeta
_{J}\left( \frac{d}{dp}-\frac{J+1}{p}+i\sqrt{\frac{\eta }{\alpha }}\frac{%
\left( 1-\lambda \theta ^{\ast }\right) }{\hbar \theta ^{\ast }}\frac{p}{%
1-\eta p^{2}}\right) F_{0}=-mc^{2}H_{+1},  \label{59}
\end{equation}%
\begin{equation}
\hbar c\theta ^{\ast }\sqrt{\frac{\alpha }{\eta }}\sqrt{1-\eta p^{2}}\xi
_{J}\left( \frac{d}{dp}+\frac{J}{p}+i\sqrt{\frac{\eta }{\alpha }}\frac{%
\left( 1-\lambda \theta ^{\ast }\right) }{\hbar \theta ^{\ast }}\frac{p}{%
1-\eta p^{2}}\right) F_{0}=-mc^{2}H_{-1}.  \label{60}
\end{equation}%
Inserting now (\ref{58}), (\ref{59}) and (\ref{60}) in (\ref{57}), the
homogeneous second-order differential equation for the DKP harmonic
oscillator is obtained as 
\begin{equation}
\left[ \left( 1-\eta p^{2}\right) \frac{d^{2}}{dp^{2}}-\eta p\frac{d}{dp}-%
\frac{J\left( J+1\right) }{p^{2}}\left( 1-\eta p^{2}\right) -\frac{\frac{%
\eta \mathcal{M}\Omega }{\hbar }\left( \frac{\mathcal{M}\Omega }{\hbar }%
-1\right) }{\left( 1-\eta p^{2}\right) }+\eta \left( \frac{\mathcal{E}^{2}}{%
\hbar ^{2}c^{2}}-\frac{\mathcal{M}^{2}c^{2}}{\hbar ^{2}}+\frac{\mathcal{M}%
^{2}\Omega ^{2}}{\hbar ^{2}}\right) \right] F_{0}=0.
\end{equation}%
After a change of variable $\beta p^{2}=\rho $, the above equation takes the
form of a hypergeometric differential equation, 
\begin{equation}
\left[ \rho \left( 1-\rho \right) \frac{d^{2}}{d\rho ^{2}}+\left( \frac{1}{2}%
-\rho \right) \frac{d}{d\rho }-\frac{J\left( J+1\right) }{4\rho }-\frac{%
\frac{\mathcal{M}\Omega }{\hbar }\left( \frac{\mathcal{M}\Omega }{\hbar }%
-1\right) }{4\left( 1-\rho \right) }+\frac{1}{4}\left( \frac{\mathcal{E}^{2}%
}{\hbar ^{2}c^{2}}-\frac{\mathcal{M}^{2}c^{2}}{\hbar ^{2}}+\frac{\mathcal{M}%
^{2}\Omega ^{2}}{\hbar ^{2}}+J\left( J+1\right) \right) \right] F_{0}=0,
\end{equation}%
whose solution can be written in term of hypergeometric functions. The
physical solution reads 
\begin{equation}
F_{0}=\mathbf{C}\text{ }\left( 1-\rho \right) ^{\frac{\mathcal{M}\Omega }{%
2\hbar }}\rho ^{\frac{J+1}{2}}\mathbf{F}\left( a^{\prime },b^{\prime },J+%
\frac{3}{2};\rho \right) ,
\end{equation}%
where%
\begin{eqnarray}
a^{\prime } &=&\frac{J+\mathcal{M}\Omega +1}{2}+\frac{1}{2}\sqrt{J\left(
J+1\right) +\frac{\mathcal{M}^{2}\Omega ^{2}}{\hbar ^{2}}-\frac{\mathcal{M}%
^{2}c^{2}}{\hbar ^{2}}+\frac{\mathcal{E}^{2}}{\hbar ^{2}c^{2}}}, \\
b^{\prime } &=&\frac{J+\mathcal{M}\Omega +1}{2}-\frac{1}{2}\sqrt{J\left(
J+1\right) +\frac{\mathcal{M}^{2}\Omega ^{2}}{\hbar ^{2}}-\frac{\mathcal{M}%
^{2}c^{2}}{\hbar ^{2}}+\frac{\mathcal{E}^{2}}{\hbar ^{2}c^{2}}}.
\end{eqnarray}%
Then, we employ the quantization condition, namely $b^{\prime }=-n$, 
\begin{equation}
\frac{J+\mathcal{M}\Omega +1}{2}-\frac{1}{2}\sqrt{J\left( J+1\right) +\frac{%
\mathcal{M}^{2}\Omega ^{2}}{\hbar ^{2}}-\frac{\mathcal{M}^{2}c^{2}}{\hbar
^{2}}+\frac{\mathcal{E}^{2}}{\hbar ^{2}c^{2}}}=-n,
\end{equation}%
where $n=0;1;...$. After a straightforward calculation, the energy spectrum
of the system is given by 
\begin{equation}
E_{n,J}=\pm c\sqrt{2m\omega \hbar \left( 2n\allowbreak +J+1\right)
+m^{2}c^{2}+\left[ 4n^{2}+4n\left( J+1\right) +J+1\right] \hbar ^{2}\left(
\alpha +\eta m^{2}\omega ^{2}\right) }.  \label{67}
\end{equation}%
However, if we remove the deformation of the space by setting $\alpha =\eta
\rightarrow 0,$ the energy spectrum (\ref{67}) becomes%
\begin{equation}
E_{n,J}=\pm c\sqrt{2m\omega \hbar \left( 2n\allowbreak +J+1\right)
+m^{2}c^{2}}.
\end{equation}%
which is in accordance with \cite{49}.

\begin{figure}[tbp]
\centering
\includegraphics[scale=0.4]{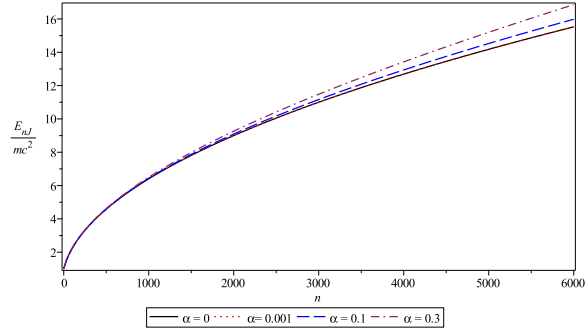}
\caption{($\frac{E_{nJ}}{mc^{2}}$ vs. $n$) for some $\protect \alpha $ values
($\protect \eta =0$). }
\label{fig:fig1}
\end{figure}

\begin{figure}[tbp]
\centering
\includegraphics[scale=0.4]{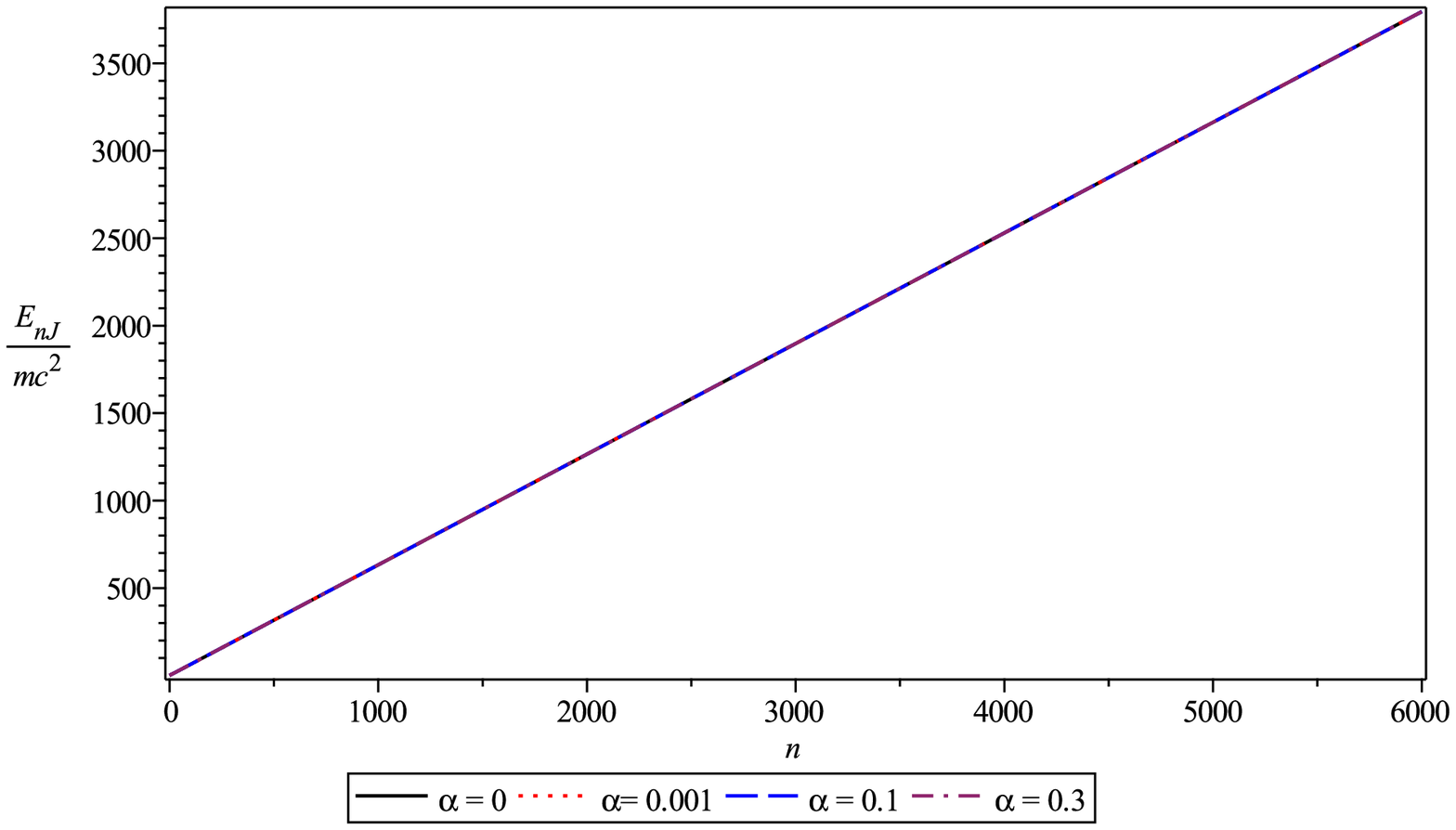}
\caption{($\frac{E_{nJ}}{mc^{2}}$ vs. $n$) for some $\protect \alpha $ values
($\protect \eta =0.001$). }
\label{fig:fig2}
\end{figure}

\begin{figure}[tbp]
\centering
\includegraphics[scale=0.4]{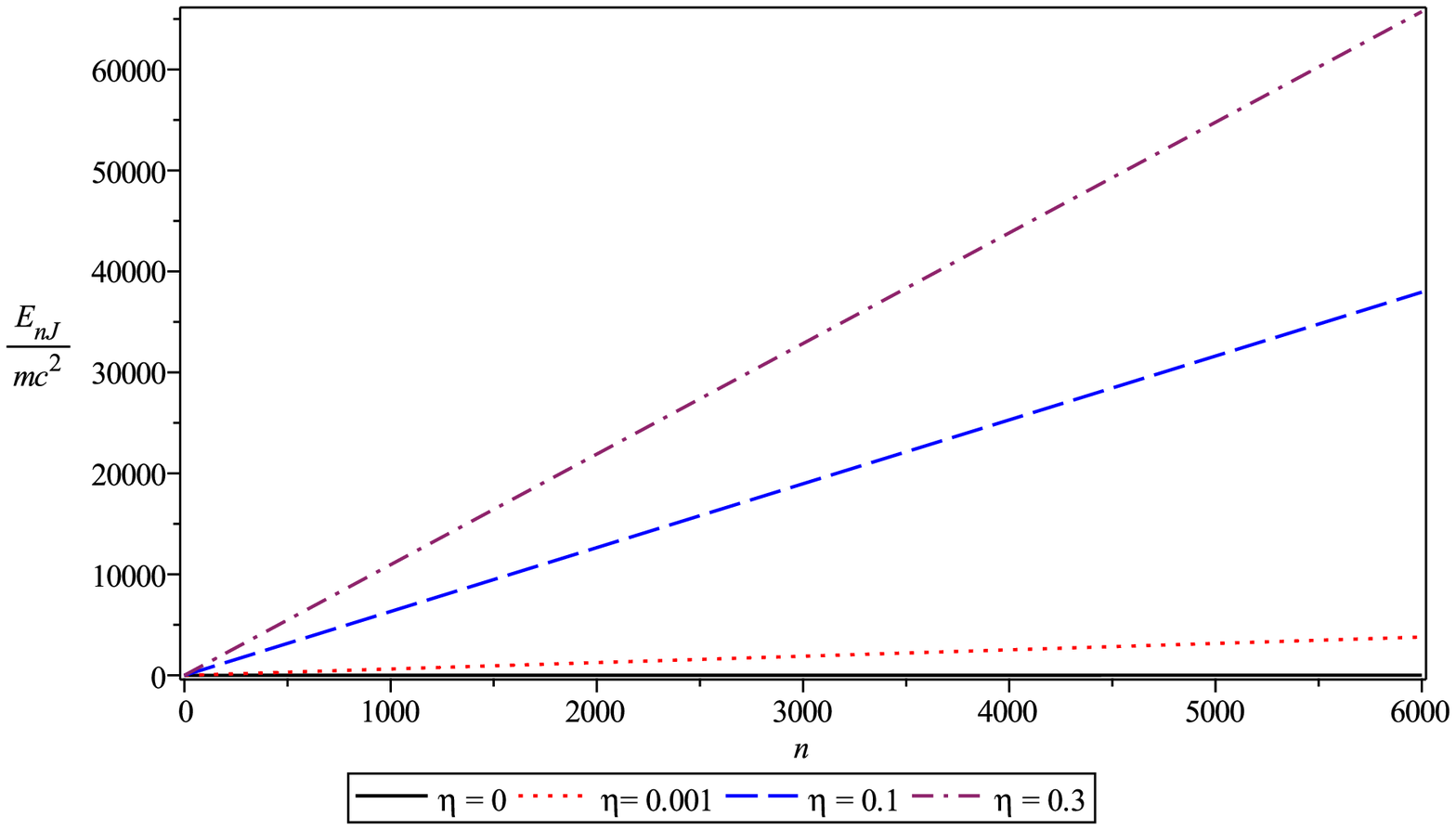}
\caption{($\frac{E_{nJ}}{mc^{2}}$ vs. $n$) for some $\protect \eta $ values ($%
\protect \alpha =0$). }
\label{fig:fig3}
\end{figure}

\begin{figure}[tbp]
\centering
\includegraphics[scale=0.4]{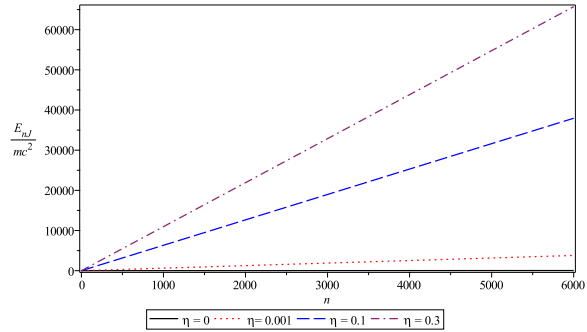}
\caption{($\frac{E_{nJ}}{mc^{2}}$ vs. $n$) for some $\protect \eta $ values ($%
\protect \alpha =10$). }
\label{fig:fig4}
\end{figure}

For the unnatural parity states, we have a coupled system, 
\begin{equation}
\hbar \theta ^{\ast }c\sqrt{\frac{\alpha }{\eta }}\sqrt{1-\eta p^{2}}\left[ 
\begin{array}{c}
\xi _{J}\left( \frac{d}{dp}+\frac{J+1}{p}+i\sqrt{\frac{\eta }{\alpha }}\frac{%
\left( 1-\lambda \theta ^{\ast }\right) }{\hbar \theta ^{\ast }}\frac{p}{%
1-\eta p^{2}}\right) G_{+1} \\ 
\zeta _{J}\left( \frac{d}{dp}-\frac{J}{p}+i\sqrt{\frac{\eta }{\alpha }}\frac{%
\left( 1-\lambda \theta ^{\ast }\right) }{\hbar \theta ^{\ast }}\frac{p}{%
1-\eta p^{2}}\right) G_{-1}%
\end{array}%
\right] +mc^{2}\Phi _{0}=0,  \label{69}
\end{equation}%
\begin{equation}
\hbar \theta c\sqrt{\frac{\alpha }{\eta }}\zeta _{J}\sqrt{1-\eta p^{2}}%
\left( \frac{d}{dp}-\frac{J+1}{p}+i\sqrt{\frac{\eta }{\alpha }}\frac{\left(
1-\lambda \theta \right) }{\hbar \theta }\frac{p}{1-\eta p^{2}}\right)
H_{0}+mc^{2}F_{+1}=EG_{+1},  \label{70}
\end{equation}%
\begin{equation}
\hbar \theta c\sqrt{\frac{\alpha }{\eta }}\xi _{J}\sqrt{1-\eta p^{2}}\left( 
\frac{d}{dp}+\frac{J}{p}+i\sqrt{\frac{\eta }{\alpha }}\frac{\left( 1-\lambda
\theta \right) }{\hbar \theta }\frac{p}{1-\eta p^{2}}\right)
H_{0}+mc^{2}F_{-1}=EG_{-1},  \label{71}
\end{equation}%
\begin{equation}
\hbar \theta c\sqrt{\frac{\alpha }{\eta }}\xi _{J}\sqrt{1-\eta p^{2}}\left( 
\frac{d}{dp}-\frac{J+1}{p}+i\sqrt{\frac{\eta }{\alpha }}\frac{\left(
1-\lambda \theta \right) }{\hbar \theta }\frac{p}{1-\eta p^{2}}\right) \Phi
_{0}+mc^{2}G_{+1}=EF_{+1},  \label{72}
\end{equation}%
\begin{equation}
\hbar \theta c\sqrt{\frac{\alpha }{\eta }}\zeta _{J}\sqrt{1-\eta p^{2}}%
\left( \frac{d}{dp}+\frac{J}{p}+i\sqrt{\frac{\eta }{\alpha }}\frac{\left(
1-\lambda \theta \right) }{\hbar \theta }\frac{p}{1-\eta p^{2}}\right) \Phi
_{0}-mc^{2}G_{-1}=-EF_{-1},  \label{73}
\end{equation}%
\begin{equation}
\hbar \theta ^{\ast }c\sqrt{\frac{\alpha }{\eta }}\sqrt{1-\eta p^{2}}\left[ 
\begin{array}{c}
\zeta _{J}\left( \frac{d}{dp}+\frac{J+1}{p}+i\sqrt{\frac{\eta }{\alpha }}%
\frac{\left( 1-\lambda \theta ^{\ast }\right) }{\hbar \theta ^{\ast }}\frac{p%
}{1-\eta p^{2}}\right) F_{+1} \\ 
+\xi _{J}\left( \frac{d}{dp}-\frac{J}{p}+i\sqrt{\frac{\eta }{\alpha }}\frac{%
\left( 1-\lambda \theta ^{+}\right) }{\hbar \theta ^{+}}\frac{p}{1-\eta p^{2}%
}\right) F_{-1}%
\end{array}%
\right] +mc^{2}H_{0}=0.  \label{74}
\end{equation}%
To proceed with the exact solution of the coupled equations associated with
the $\left( -1\right) ^{J+1}$ parity states, Eqs. (\ref{69}), (\ref{70}), (%
\ref{71}) and (\ref{72}), (\ref{73}), (\ref{74}) for the $J=0$ case are
transformed into 
\begin{eqnarray}
\left( 
\begin{array}{c}
F_{+1} \\ 
G_{+1}%
\end{array}%
\right) &=&\frac{\hbar \theta c\sqrt{\frac{\alpha }{\eta }}}{E^{2}-m^{2}c^{4}%
}\sqrt{1-\eta p^{2}}\left( \frac{d}{dp}-\frac{1}{p}+i\sqrt{\frac{\eta }{%
\alpha }}\frac{\left( 1-\lambda \theta \right) }{\hbar \theta }\frac{p}{%
1-\eta p^{2}}\right) \left( 
\begin{array}{cc}
E\xi _{J} & 0 \\ 
mc^{2}\xi _{J} & 0%
\end{array}%
\right) \left( 
\begin{array}{c}
\Phi _{0} \\ 
H_{0}%
\end{array}%
\right) , \\
\left( 
\begin{array}{c}
F_{-1} \\ 
G_{-1}%
\end{array}%
\right) &=&\frac{\hbar \theta c\sqrt{\frac{\alpha }{\eta }}}{E^{2}-m^{2}c^{4}%
}\sqrt{1-\eta p^{2}}\left( \frac{d}{dp}+i\sqrt{\frac{\eta }{\alpha }}\frac{%
\left( 1-\lambda \theta \right) }{\hbar \theta }\frac{p}{1-\eta p^{2}}%
\right) \left( 
\begin{array}{cc}
0 & mc^{2}\xi _{J} \\ 
0 & E\xi _{J}%
\end{array}%
\right) \left( 
\begin{array}{c}
\Phi _{0} \\ 
H_{0}%
\end{array}%
\right) ,
\end{eqnarray}%
\begin{eqnarray}
\left[ \left( 1-\eta p^{2}\right) \frac{d^{2}}{dp^{2}}-\eta p\frac{d}{dp}-%
\frac{\frac{\mathcal{M}\Omega \eta }{\hbar }}{\left( 1-\eta p^{2}\right) }-%
\frac{\eta \mathcal{M}^{2}\Omega ^{2}}{\hbar ^{2}}\frac{\eta p^{2}}{\left(
1-\eta p^{2}\right) }+\eta \frac{\mathcal{E}^{2}-\mathcal{M}^{2}c^{4}}{\hbar
^{2}c^{2}}-2\eta \left( \frac{\mathcal{M}\Omega }{\hbar }-\frac{1}{2}\right) %
\right] \Phi _{0} &=&0, \\
\left[ \left( 1-\eta p^{2}\right) \frac{d^{2}}{dp^{2}}-\beta \eta \frac{d}{dp%
}-\frac{\frac{\mathcal{M}\Omega \eta }{\hbar }}{\left( 1-\eta p^{2}\right) }-%
\frac{\beta \mathcal{M}^{2}\Omega ^{2}}{\hbar ^{2}}\frac{\eta p^{2}}{\left(
1-\eta p^{2}\right) }+\eta \frac{\mathcal{E}^{2}-\mathcal{M}^{2}c^{4}}{\hbar
^{2}c^{2}}\right] H_{0} &=&0.
\end{eqnarray}%
Following the same method given in the spin-0 case, the energy spectra of
the vector DKP oscillator in curved Snyder space are given by%
\begin{eqnarray}
E_{\Phi _{0}} &=&\pm c\sqrt{2\hbar m\omega \left( 2n+2\right)
+m^{2}c^{2}+\left( 4n^{2}+4n\right) \hbar ^{2}\left( \alpha +\eta
m^{2}\omega ^{2}\right) }, \\
E_{H_{0}} &=&\pm c\sqrt{2\hbar m\omega \left( 2n+1\right) +m^{2}c^{2}+\left(
4n^{2}+4n+1\right) \hbar ^{2}\left( \alpha +\eta m^{2}\omega ^{2}\right) }.
\end{eqnarray}%
Finally, we will plot the natural-parity $E_{n,J}$ levels as given in Eq. (%
\ref{67}) as a function of variable $n$ by taking $\hbar \omega =10MeV$ and $%
mc^{2}=1GeV$. We will also take $J=0$. We will first make the energy
function dimensionless for a better analysis, namely, 
\begin{equation*}
\frac{E_{n,J}}{mc^{2}}=\pm \sqrt{2\frac{\hbar \omega }{mc^{2}}\left(
2n\allowbreak +J+1\right) +1+\left[ 4n^{2}+4n\left( J+1\right) +J+1\right]
\left( \alpha \frac{\hbar ^{2}}{m^{2}c^{2}}+\eta \frac{\hbar ^{2}\omega ^{2}%
}{c^{2}}\right) }.
\end{equation*}%
We will work only for the positive values of $\frac{E_{n,J}}{mc^{2}}$. We
should note that, for these numerical values and for $\hbar =c=1$, the
coefficient of the $\alpha $-parameter ($\frac{\hbar ^{2}}{m^{2}c^{2}}$) is
of the order $10^{-6}$ and the coefficient of the $\eta $-parameter ($\frac{%
\hbar ^{2}\omega ^{2}}{c^{2}}$) is of the order $10^{2}$. Generally, the
contribution of the $\alpha $-parameter increases with $m^{-2}$ and the
contribution of the $\eta $ parameter increases with $\omega ^{2}$ as the $%
\frac{\hbar ^{2}}{c^{2}}$ factor is common. This fact can be seen by
comparing the first set of the figures (\ref{fig:fig1}), (\ref{fig:fig2})
with the second set given by Figs. (\ref{fig:fig3}), (\ref{fig:fig4}).

In Fig. (\ref{fig:fig1}), we fix $\eta =0$ and plot the energy eigenvalues
for different values of the parameter $\alpha $. The $\alpha $-dependence
becomes clearer as $n$ increases. If we change $\eta $ to a non-zero value,
namely $\eta =0.001$ as in Fig. (\ref{fig:fig2}), we see that the behavior
of the curves changes and the $\alpha $-dependence becomes negligible as the
non-zero $\eta $ value dominates the behavior. In Figs. (\ref{fig:fig3}) and
(\ref{fig:fig4}), we fix $\alpha =0$ and $\alpha =10$, respectively. We plot
for different values of the parameter $\eta $. It can be seen that changes
in the $\eta $ values affect the behavior of the energy spectrum
extensively. However, the $\alpha $-dependence is ignorable as its
coefficient is very small with respect to the $\eta $-term as mentioned
above. 

\section{Conclusion}

We studied the three dimensional scalar and vector DKP oscillators in
momentum space for the case of Snyder-de Sitter model, which is an extension
of the Snyder model which is called the triply special relativity (TSR).

By introducing the technique of vector spherical harmonics, we obtained the
exact energy spectrum and corresponding eigenfunctions expressed in terms of
Jacobi polynomials for the both cases. We employed the condition that yields
a polynomial solution for a hypergeometric function as the quantization
condition to obtain the energy spectra. Furthermore, we argued that the $%
\eta $-dependence is more significant for the numerical values of a typical
physical system. For $\eta=0$, the behavior of the curves is governed by the
parameter $\alpha$, especially for large $n$ values. However, even for a
small non-zero $\eta$, the $\alpha$-dependence becomes negligible. This fact
is also observed when we fix $\alpha$ and plot for different values of the
parameter $\eta$.

The energy levels show a dependence on $n^{2}$ which explains the
confinement at the high energy sector. It should be emphasized that in the
limit $\alpha \rightarrow 0$ one recovers the Snyder algebra and, in the
limit $\eta \rightarrow 0$ one recovers the deformed Heisenberg algebra in
de Sitter space as expected. The undeformed Heisenberg algebra is obtained
in the limit when $\alpha $ and $\eta $ both tend to zero.

\end{document}